\documentstyle[a4,12pt,cite,psfig]{article}

\begin{document}
\begin{titlepage}
\Large

\begin{center}
Numerical Analysis of the Double Scaling Limit \\
in the IIB Matrix Model
\end{center}
\vspace{0.5cm}
\begin{center}
\large
S.Horata \footnote[1]{E-mail address: horata@ccthmail.kek.jp} and 
H.S.Egawa \footnote[2]{E-mail address: egawah@ccthmail.kek.jp}
\\
\vspace{0.3cm}
\normalsize
$^{\ast, \dag}$ 
Theory Division, Institute of Particle and Nuclear Studies, \\
KEK, High Energy Accelerator Research Organization, \\
Tsukuba, Ibaraki 305-0801, Japan \\

\vspace{0.3cm}
$^{\dag}$ 
Department of Physics, Tokai University, \\
Hiratsuka, Kanagawa 259-1292, Japan
\end{center}

\vspace{0.5cm}
\begin{abstract}
The bosonic IIB matrix model is studied using a numerical method.
This model contains the bosonic part of the IIB matrix model conjectured
to be a non-perturbative definition of the type IIB superstring theory.
The large $N$ scaling behavior of the model is shown performing a Monte
Carlo simulation.
The expectation value of the Wilson loop operator is measured and the
string tension is estimated.
The numerical results show the prescription of the double scaling limit. 

\vspace{.3cm}
{\sl PACS:} 
11.25.sq

{\sl Keywords:}
IIB Matrix theory;
Numerical analysis;
Scaling structure;
Area law 

\end{abstract}
\end{titlepage}
\section{Introduction} 
Recent works in string theory have proposed some models as
non-perturbative formulations\cite{BFSS,IKKT,AIKKTT}.
Especially, the IIB matrix model\cite{IKKT,AIKKTT} has been considered as
a constructive definition of the type IIB superstring theory.
This model is zero-volume limit of the ten-dimensional large $N$
supersymmetric Yang-Mills theory and defined by the following action,
\begin{equation}
 S = - \frac{1}{g^2} {\rm tr} \left( \frac{1}{4}[A_\mu, A_\nu]^2 
     + \frac{1}{2} \bar{\psi} \Gamma^\mu [A_\mu,\psi]\right) ,
\end{equation}
where $A_\mu$ and $\psi$ are $N \times N$ traceless Hermitian matrices. 
The interesting feature of this model is that the space-time coordinates
are considered as the eigenvalues of these matrices.
Then, we expect that the fundamental issues including the dimensionality
and the quantum gravity can be understood by studying the dynamics of
the model.

To take the continuum limit ($N \rightarrow \infty$) for the IIB Matrix 
model, a sensible double scaling limit should be determined dynamically.
The scaling property of the model has been studied with the light-cone
string field Hamiltonian of the type IIB superstring theory\cite{FKKT}.
The scaling property of the two important quantities of the model,
the string scale ($\alpha'$) and the string coupling constant
($g_{str}$) are determined as follows,
\begin{eqnarray}
 \alpha' &\sim& g^2 N^{a+b} = g^2 N^{\gamma} \sim Constant, \nonumber \\
 g_{str} &\sim& N^{a-b}.
\end{eqnarray}
For the finite value of the string coupling constant ($g_{str}$), one
have a restriction, $a=b$.
In the IIB matrix model, the exponent ($\gamma$) plays an important
role that one can take the large $N$ limit as the continuum limit for
the IIB Matrix model.
This exponent is determined dynamically.

For studying the dynamical aspects of the model, some Monte Carlo
simulations have been performed.
In Ref.\cite{KNS,KS}, the existence of the large $N$ limit of bosonic
$SU(N)$ Yang-Mills matrix model for $D>2$ has been discussed analytically.  
Then, in Ref.\cite{HNT}, the bosonic model has been studied with a
analytical method, the $1/D$ expansion, as well as a numerical one.
The numerical results support the $1/D$ expansion as an effective tool
detecting the large $N$ scaling behavior of the model.
The leading term of the $1/D$ expansion at $D>3$ suggests that the
exponent ($\gamma$) takes a value of 1.
We study the model in ten dimensions and reconfirm the expected scaling
behavior by a numerical method with a larger size matrix.

In analogy with the two-dimensional model, the Eguchi-Kawai model, 
we study the scaling property of the Wilson loop in ten-dimensions. 
The area law of the Wilson loop operator has been found in the
four-dimensional model in Ref.\cite{AABHN}.
In this article, we also obtain that the area dependence of the Wilson
loop obeys the area law in ten dimensions.
We calculate the string tension from the area law of the Wilson loop.
We thus consider that the scaling property of the string tension is
estimated in ten-dimensional model.

This paper is organized as follows. In section 2, we review the model
and some perturbative analysis. In section 3, we show the numerical
results and the scaling property of the model. Then, we present the data 
which show the existence of the double scaling limit in the
model. Finally, in section 4, we summarize and discuss our numerical
results.

\section{Large $N$ behavior of correlation functions}
First, let us remind the perturbative arguments of the bosonic model and 
describe briefly the large $N$ behavior of the correlation
functions of the gauge fields based on \cite{HNT}.

The bosonic model of the IIB matrix model is given by
\begin{equation}
 S_{bosonic} = -\frac{1}{g^2} {\rm tr} [A_\mu, A_\nu][A^\mu,A^\nu], \label{action}
\end{equation}
where $A_\mu$ are $N \times N$ Hermitian matrices representing the
ten-dimensional gauge fields.
The coupling constant ($g$) is nothing but a scale parameter and is
absorbed with the rescaling of the gauge field ($A_\mu$) as $A_\mu
\rightarrow \frac{1}{\sqrt{g}}A_\mu$.

The Schwinger-Dyson equation is given by
\begin{equation}
 0 = \int dA \frac{\partial}{\partial A} \left( {\rm tr}\left( A_\mu  e^{-S_{bosonic}} \right)\right),
\end{equation}
leading to the relation of the correlation function,
\begin{equation}
-<{\rm tr}([A_\mu,A_\nu]^2)> = D(N^2 - 1) g^2. \label{trAA-SD}
\end{equation}

For the estimation of the large $N$ behavior of the correlation
function, $<{\rm tr}([A_\mu,A_\nu]^2)>$, the matrices ($A_\mu$) are
decomposed into the diagonal parts ($X_\mu^i$) and the off diagonal parts
($\tilde{A_\mu}^{ij}$), and eq.(\ref{trAA-SD}) is taken up to the second
order of the off diagonal elements ($\tilde{A_\mu}^{ij}$),
\begin{equation}
 <{\rm tr}[A_\mu,A_\nu]^2> \simeq 
  2 { <{\rm tr}[X_\mu, \tilde{A}_\nu] [X_\mu, \tilde{A}_\nu]>
  - < {\rm tr}[X_\mu, \tilde{A}_\nu] [X_\nu, \tilde{A}_\mu]> }.
\end{equation}
Counting the order of the diagonal parts, the large $N$ behavior of the
leading term of the correlation function counting with the order of the
diagonal parts is given by
\begin{equation}
 <{\rm tr}[A_\mu,A_\nu]^2> \sim g^2 N^2.
\end{equation}
For the finite value of the correlation function, an upper limit of the
typical scale of the extend of the space-time, for example can be $R^2 =
<{\rm tr}A^2>$, is suggested as $R^2  \leq N^{1/2}$ perturbatively and
from the $1/D$ expansion the large $N$ behavior is shown as\cite{HNT}
\begin{equation}
 R^2 \sim g N^{1/2}.
\end{equation}
In the similar manner the correlation functions can be calculated
perturbatively.

Next, we consider the Wilson loop operator in the IIB matrix model.
The Wilson loop operator and the large $N$ behavior have been studied 
with the light-cone string field theory of the type IIB
superstring\cite{FKKT}.
The Wilson loop operator ($w(C)$) is defined as,
\begin{equation}
 w(C) = {\rm tr}(v(C)),
\end{equation}
where $C$ denotes the closed path and $v(C)$ is defined as $v(C) = U_\mu
\cdots U_\mu =  P_C \exp (i \oint d \sigma k_\mu A_\mu) $ in the
bosonic model. 
The matrices ($U_\mu$) are considered as the unitary matrices,
\begin{equation}
 U_\mu = \exp ( i \int d l A_\mu).
\end{equation}

In the ordinary lattice gauge field theory, the expectation value of the
Wilson loop operator which spreads a large area behaves as follows,
\begin{equation}
 w(I,J) \sim \exp (- K I\times J),
\end{equation}
where $I$ and $J$ are the side lengths of the rectangular loop and $K$
denotes the string tension.
In the same analogy, we study the Wilson loop operator of the bosonic
model of the IIB matrix model.

From the scaling relation of two-dimensional Eguchi-Kawai model,
\begin{equation}
 g^2 N \sim Constant,
\end{equation}
It is expected that the similar scaling relation also holds in the IIB matrix
model as\cite{FKKT}
\begin{equation}
 \alpha' \sim g^2 N^{\gamma} \sim Constant. \label{IIB-SCALING}
\end{equation}
The exponent ($\gamma$) should be determined dynamically from
the model.
For the large $N$ limit, the parameter ($g^2 N^{\gamma}$) must be fixed
in the IIB matrix model in the same manners as the parameter ($g^2 N$) must
be fixed in the Eguchi-Kawai model.

We notice that the bosonic model is equivalent to the $D>2$ Eguchi-Kawai
model in the weak coupling limit.
Since the $U(1)^D$ symmetry rotates all the eigenvalues by the same
angle, the following expansion is valid in the weak coupling region,
\begin{equation}
 U_\mu \sim e^{i \alpha_\mu} e^{i A_\mu},
\end{equation}
where $\alpha_\mu$ take constant values due to the $U(1)^D$ symmetry and
$A_\mu$ are small.
The bosonic model action can be obtained by expanding the action of 
the ten-dimensional Eguchi-Kawai model in terms of $A_\mu$.
When the higer order terms of $A_\mu$ can be neglected, we can obtain
the area law of the Wilson loop operator as
\begin{equation}
 w(I \times J) = <{\rm tr}(U_\mu \cdots U_\mu)_{I \times J}> \simeq 
  <{\rm tr}(e^{i A_\mu} \cdots e^{i A_\mu})_{I \times J}> \sim \exp ( -K (I \times J)). \label{IIB-AREALAW}
\end{equation}
In Ref.\cite{AABHN}, the area dependence of the Wilson loop operator has 
been measured and found the area law in $D=4$.

In following section, we will show that the area law holds in the bosonic
model of the IIB matrix model in $D=10$.

\section{Monte Carlo simulation and Large N behavior of the bosonic model}
For numerical simulation of the model, we consider the partition
function,
\begin{equation}
 Z = \int dA e^{-S_{bosonic}},
\end{equation}
where the action $S_{bosonic}$ is given by eq.(\ref{action}) and the
measure of gauge fields is defined by
\begin{equation}
 dA = \prod_{\mu=1}^{10} \left[ \prod_{i=1}^{N} \prod_{j=i}^{N}
 d(A_{\mu}^{ij}) \right].
\end{equation}

The action is quadratic with respect to each component, which means that
we can update each component by generating gaussian random number in the
heat-bath and the Metropolis algorithm.

\begin{figure}[t]
\vspace{-10mm}
\centerline{\psfig{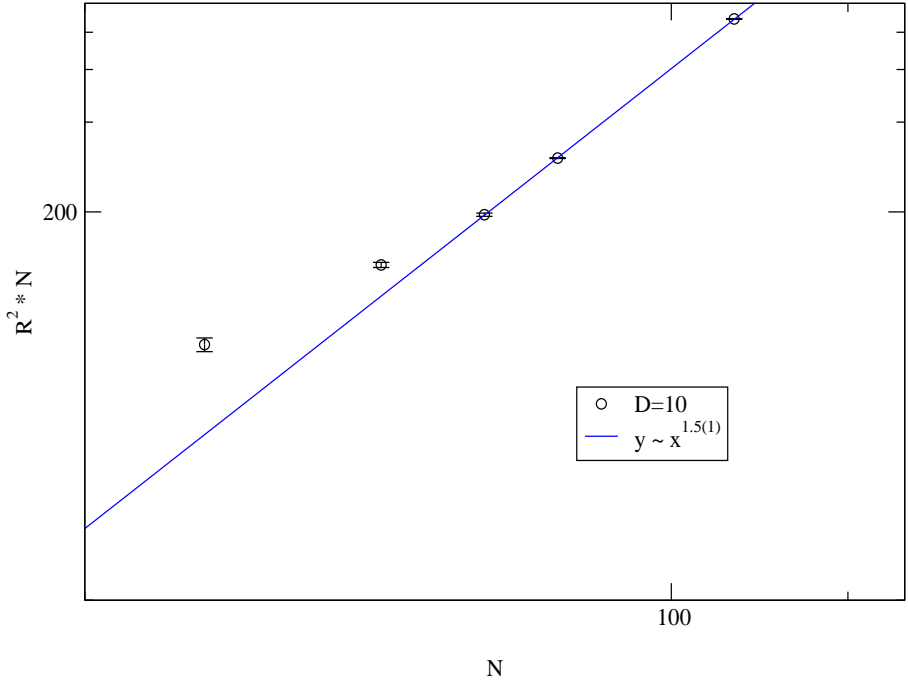}}
\vspace{3mm}
\caption
{
The measurement results of the extent of the space time.
}
\label{fig:AA}
\end{figure}

\begin{figure}[t]
\centerline{\psfig{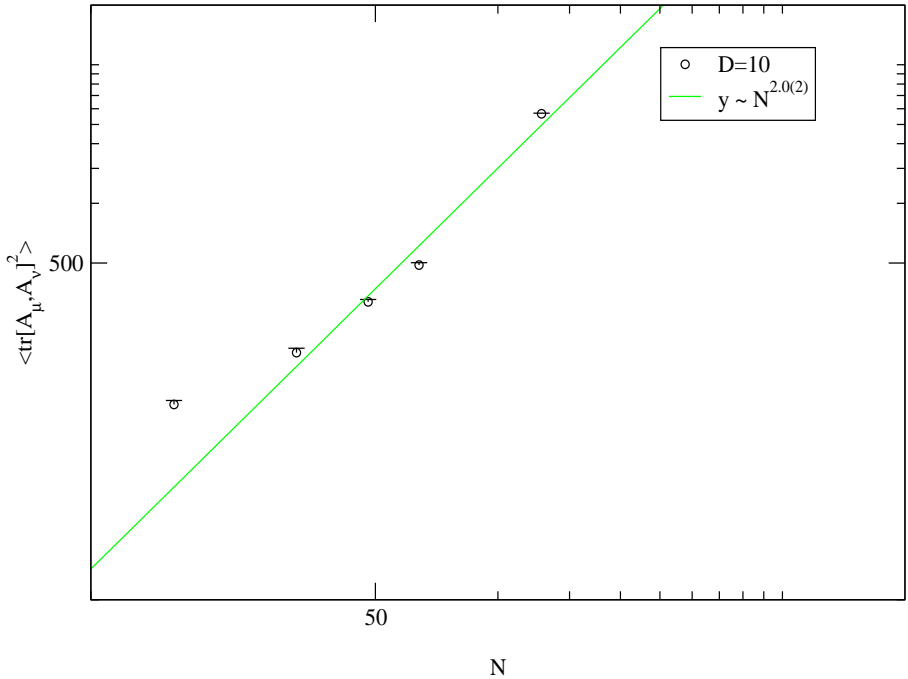}}
\caption
{
The measurement results of the extent of the correlation function of ($A_\mu$).
}
\label{fig:FF}
\end{figure}

To confirm the large $N$ behavior of the model, we measure the following
expectation values,
\begin{eqnarray}
 R^2 = <\frac{1}{N} {\rm tr} (A^2) >, \nonumber \\
 < \frac{1}{N} {\rm tr} ([A_\mu, A_\nu]^2) >.
\end{eqnarray}
From the perturbative calculation, the large $N$ behaviors are shown as
\begin{eqnarray}
& & R^2 \sim g N^{1/2}, \nonumber \\
& & < \frac{1}{N} {\rm tr} ([A_\mu, A_\nu]^2) > \sim g^2 N.
\end{eqnarray}
In Fig.\ref{fig:AA} and Fig.\ref{fig:FF}, we show the numerical
results of the extent of space-time ($R^2$) and the correlation function
($<{\rm tr}([A_\mu, A_\nu])^2>$) for $D=10$ with $N = 16, 32, 48, 64,
128$, respectively.
We obtain the large $N$ behavior as
\begin{eqnarray}
& & R^2 \sim g N^{0.5(1)}, \nonumber \\
& & < \frac{1}{N} {\rm tr} ([A_\mu, A_\nu]^2) > \sim g^{2} N^{-1.0(1))}.
\end{eqnarray}
By the simulation using the larger size matrix, the numerical result
get close to the expected results.
%
\vspace{10mm}

\begin{figure}[t]
\centerline{\psfig{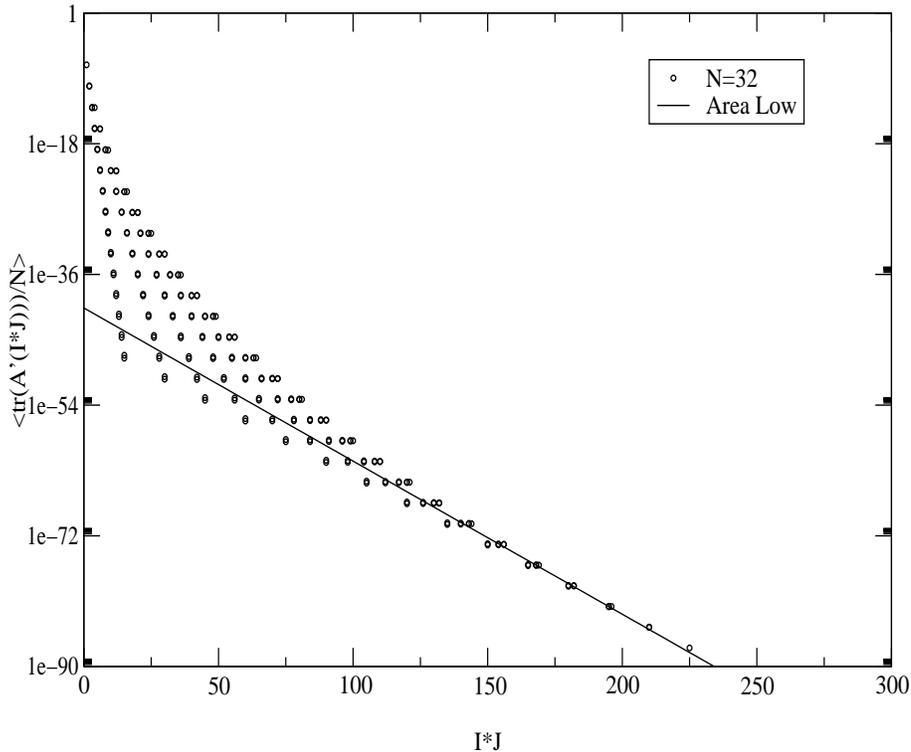}}
\caption
{
The large area behavior of the Wilson loop operator
 $w(I \times J)$. Each dot corresponds to different choice of $I$ and $J$.
}
\label{fig:arealaw}
\end{figure}

Then, we consider the Wilson loop operator,
\begin{equation}
 w(C) = <\frac{1}{N} {\rm tr} (e^{i A_\mu} \cdots e^{i A_\nu}) >.
\end{equation}
We take the loop ($C$) as the rectangular ($I \times J$) where we select 
any two direction ($\mu,\nu$) in ten dimensions. 

We show the measurement results of the loop operator in
Fig.\ref{fig:arealaw}. 
Since the dependence of the direction of the rectangular is not found,
we consider that in the bosonic model the isotropy of the
ten-dimensional space-time is not broken down spontaneously.

Then, we calculate the large $N$ behavior by the Wilson loop operator.
From the numerical results, the Wilson loop operator, $<\frac{1}{N} {\rm
tr} (e^{i A_\mu} \cdots e^{i A_\nu})_{I \times J} >$, closes to the
exponential curve with the large size area.
It means that the Wilson loop operator obeys the area law
eq.(\ref{IIB-AREALAW}).

\begin{figure}[t]
\centerline{\psfig{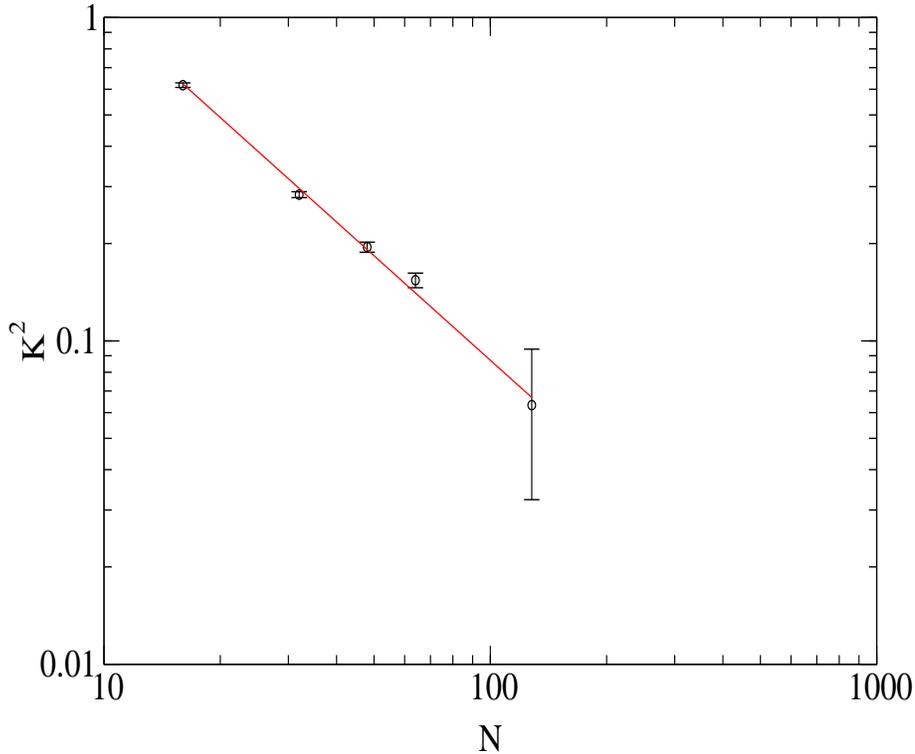}}
\caption
{
The measurement results of the string tension $a$.
}
\label{fig:kappa}
\end{figure}

Then, we can obtain the string tension.
In Fig.\ref{fig:kappa}, we plot the string tension ($K$).
%

%

We find the large $N$ behavior of the string tension ($K$),
\begin{equation}
 K \sim g^{-2} N^{-1.07(1)}.
\end{equation}
From the string tension ($K$), the string scale ($\alpha'$) is estimated 
as 
\begin{equation}
 \alpha' \sim 1/K \sim g^{2} N^{1.07(1)} = Constant. \nonumber
\end{equation}
We remark that the string field theory of the light-cone frame
suggests\cite{FKKT}
\begin{equation}
 \alpha ' \sim g^{2} N = Constant.
\end{equation}
We consider that the numerical result closes to the analytic one and
that the four-dimensional model has the same scaling
property\cite{AABHN}.

Furthermore the numerical result suggests that the large $N$ behavior of
the square root of the string tension approximately equal to the inverse of the
extent of the space-time.
\begin{equation}
 K^{1/2} \sim g^{-1} N^{-0.54(1)} \sim R^{-1}.
\end{equation}
It means that the Planck scale of the theory has the same scaling
property of the extent of the space time in ten-dimensions.
It also holds on the two-dimensional model\cite{NN} and the
four-dimensional model\cite{AABHN}.

\section{Summary and Discussion}
%
Let us summarize the main points made in our calculation.
We confirm that the Wilson loop operator in the ten-dimensional bosonic
model obeys the area law similar to the two and four-dimensional
model\cite{NN,AABHN}.
For the scaling behavior of the bosonic model of the IIB matrix model,
our numerical estimation is
\begin{equation}
 \alpha' \sim g^{2} N^{1.07(1)} = Constant.
\end{equation}

Our results show the ten-dimensional space-time extends with $g^{1/2}
N^{0.25(10)}$ and the extent scale of the space-time approximately equal
to the Planck scale ($l$), $l \sim K^{-1/2}$.
Furthermore, these results are consistent to the suggestion from 
the string field theory on light-cone frame\cite{FKKT} and the $1/D$
expansion\cite{HNT}.
From the numerical results, we consider that the ten-dimensional
bosonic model and the four-dimensional model have the same scaling
property.
In this article, we calculate only the bosonic model of the IIB matrix
model.
For the future work, we are also considering the numerical simulation
of the full model including the fermionic term.
The supersymmetric four-dimensional model has been studied in Ref.\cite{AABHN}.
Optimistically, we expect that we can simplify the fermionic term
with the perturbative calculation.
In Ref.\cite{AIKKT}, it is claimed that the model with the 1-loop effective
action of the IIB matrix model produces the four-dimensional space-time from
the ten-dimensional space-time.
We thus make preparations the calculation of the modified model
including the fermionic term.
The improved supersymmetric model including the fermionic term is
studied\cite{AABHN2}.
\begin{center}
{\Large Acknowledgements}
\end{center}
We would like to thank T.Yukawa, N.Ishibashi, Y.Kitazawa and H.Kawai.
Furthermore, we are grateful to F.Sugino, N.Tsuda, S.Oda and especially
J.Nishimura for fruitful discussions and advice. We are also grateful to
the members of the KEK theory group. Numerical calculations were
performed using the NEC SX4 (Tokai University) and the originally
designed cluster computer for quantum gravity and strings, CCGS (KEK).

\end{document}